\documentclass[graybox]{svmult}

\usepackage{mathptmx}       
\usepackage{helvet}         
\usepackage{courier}        
\usepackage{type1cm}        
\usepackage{makeidx}         
\usepackage{graphicx}        
\usepackage{multicol}        
\usepackage[bottom]{footmisc}

\makeindex             

\def\hi{\hangindent\parindent\hangafter=1\noindent}

\begin{document}

\title*{On the future of astrostatistics:
statistical foundations and statistical practice}
\titlerunning{The future of astrostatistics}
\author{Thomas J. Loredo}
\institute{Thomas J. Loredo \at Center for Radiophysics \& Space Research,
Cornell University, Ithaca, NY 14853-6801, \email{loredo@astro.cornell.edu}}

\maketitle


\begin{quote}\em
This paper summarizes a presentation for a panel discussion on ``The Future
of Astrostatistics'' held at the {\em Statistical Challenges in Modern
Astronomy~V} conference at Pennsylvania State University in June 2011.  I
argue that the emerging needs of astrostatistics may both motivate and benefit
from fundamental developments in statistics.  I highlight some recent work
within statistics on fundamental topics relevant to astrostatistical practice,
including the Bayesian/frequentist debate (and ideas for a synthesis),
multilevel models, and multiple testing.  As an important direction for future
work in statistics, I emphasize that astronomers need a statistical framework
that explicitly supports unfolding chains of discovery, with acquisition,
cataloging, and modeling of data not seen as isolated tasks, but rather as
parts of an ongoing, integrated sequence of analyses, with information and
uncertainty propagating forward and backward through the chain.  A
prototypical example is surveying of astronomical populations, where source
detection, demographic modeling, and the design of survey instruments and
strategies all interact.

The panel was moderated by Jogesh Babu; the other panelists were Eric
Feigelson (on the past and future of astrostatistics), Jeffrey Scargle (on
challenges and opportunities in astrostatistics), and David van Dyk (on
massive datasets and complex models).  I am indebted to Eric Feigelson for
an edited transcript of my recorded remarks, on which this paper is
based.  I limited my subsequent editing to preserve the informality
of the panel discussion setting.

Due to page limitations, the published version was abridged; it omits
Fig.~1 and the associated text.
\end{quote}





\index{Bayesian inference} \index{Bayesian inference! debate with frequentists} 
\index{False Detection Rate}

\section{The Frequentist-Bayesian debate}

The future of astrostatistics is linked to the future of statistics as a
discipline.  The emerging needs of astrostatistics may both motivate and
benefit from {\it fundamental} developments in statistics.  This is a two-way
street.

Christopher Genovese told us earlier in the conference that, within
statistics, the debate between frequentist and Bayesian approaches has largely
faded from view.  He noted that, although nontrivial philosophical and
conceptual differences certainly exist, statisticians recognize that there are
situations where each approach has an advantage, and both are used
successfully.

My outsider view of contemporary statistics supports the assessment that
debate about foundations has faded in recent years.  But I do not see this as
a positive development, and I disagree with any prescription that fundamentals
should not be seriously discussed and researched.  Issues at the foundations
of statistics are not merely philosophical.   Where one comes down on
foundational issues has significant implications for statistical practice.  I
would urge statisticians to think more rather than less about the foundations
of their discipline, and to consider doing so in closer partnership with the
scientist consumers of their methods.  Despite being an outsider to
statistics, I take this position emboldened by being in good company from
within the discipline, and by the seriousness of the topic.  For I see
statistics as a kind of theory of the scientific method---at least, that part
of the scientific method that may be described with quantitative
precision---giving all scientists a vested interest in the field's
development.



Prominent statisticians who have contributed enormously to statistical
practice continue to embrace the struggle with the foundations and
fundamentals of statistical inference.  Bradley Efron (2010), whose work
mostly adopts the frequentist approach, recently lamented the absence of
attention to foundations:
\begin{quote}
Methodology by itself is an ultimately frustrating exercise.  A little
statistical philosophy goes a long way but we have had very little in the
public forum these days.
\end{quote}
In his 2004 American Statistical Association (ASA) Presidential Address (Efron
2005), he asserted:
\begin{quote} 
The 250-year debate between Bayesians and frequentists is unusual among
philosophical arguments in actually having important practical
consequences\ldots. Broadly speaking, Bayesian statistics dominated 19th
Century statistical practice while the 20th Century was more frequentist. 
What's going to happen in the 21st Century? \ldots\ I strongly suspect that
statistics is in for a burst of new theory and methodology, and that this
burst will feature a combination of Bayesian and frequentist reasoning.
\end{quote}
Efron sees empirical Bayes methods as a promising frequentist/Bayesian
hybrid approach (Efron 2010; see the discussion accompanying his article for
critical assessments); I will have more to say about this below.


To cite another example, in his 2005 ASA President's Invited Address (Little
2006), Roderick Little wrote:
\begin{quote}
Pragmatists might argue that good statisticians can get sensible answers under
Bayes or frequentist paradigms; indeed maybe two philosophies are better than
one, since they provide more tools for the statistician's toolkit\ldots. I am
discomforted by this ``inferential schizophrenia.''  Since the Bayesian (B)
and frequentist (F) philosophies can differ even on simple problems, at some
point decisions seem needed as to which is right.  I believe our credibility
as statisticians is undermined when we cannot agree on the fundamentals of our
subject.
\end{quote}
Little, whose work has mostly adopted the Bayesian approach, has recently
tried to work out principles for best practices, pulling strengths from each
approach.  Roughly speaking, his {\em calibrated Bayes} synthesis relies on
Bayesian methods for inference under a model, but holds an important role for
frequentist ideas in model assessment.  He feels strongly that Bayesian
methods are insufficiently taught to statisticians.  But he also criticizes
advocates of Bayesian methods for not sufficiently assessing their modeling
assumptions.

With such leading lights harping on the need to examine fundamentals, why is
there so little of what one might call ``foundational self-examination'' in
statistics? Andrew Gelman (2010), in a discussion of the empirical Bayes
synthesis of Efron (2010), presents three meta-principles of statistics, among
them one shedding a bit of light on this question:
\begin{quote}
My second meta-principle of statistics is the {\it methodological attribution
problem}, which is that the many useful contributions of a good statistical
consultant, or collaborator, will often be attributed to the statistician's
methods or philosophy rather than to the artful efforts of the statistician
himself or herself.  The result is that each of us tends to come away from a
collaboration or consulting experience with the warm feeling that our methods
really work, and that they represent how scientists really think.  In stating
this, I am not trying to espouse some sort of empty pluralism\ldots.  I think
we all have to be careful about attributing too much from our collaborators'
and clients' satisfaction with our methods.
\end{quote}
The meta-principle speaks to the absence of reflection on foundations:  truly
talented statisticians adopting different approaches get good work done; the
approach they adopt seems not to matter.  But Gelman's comment about ``empty
pluralism'' is important.  Satisfaction with the current ``inferential
schizophrenia'' in statistics is not justified by past successes.  Brilliant
analysts can rely on on unarticulated intuition, but the rest of us need sound
principles, if only they can be uncovered.  (We can also benefit from
collaboration, but that's another topic!)



Susie Bayarri and James Berger provide concrete examples of methodological
advances coming from foundational research in their survey, ``The Interplay of
Bayesian and Frequentist Analysis'' (Bayarri \& Berger 2004).  They argue that
``the debate is far from over and, indeed, should continue, since there are
fundamental philosophical and pedagogical issues at stake,'' with significant
implications for practice.  They review research that combines frequentist and
Bayesian ideas resulting in new directions for statistical practice, including
work on frequentist performance of Bayesian procedures, predictive assessment
of models, conditional frequentist testing, and so forth.  To highlight just
one area with practical consequences:  Conditional frequentist testing is an
alternative to traditional hypothesis testing with $p$-values (astronomers'
``significance levels'') that I have found to be appealing to astronomers I
work with, because it does what they thought their $p$-values were doing.  It
also happens to be closely related to model comparison with Bayes factors, and
so serves as a natural bridge between Bayesian and frequentist thinking. 
Bayarri and Berger also discuss areas where the two approaches seem to
fundamentally disagree, such as multiple testing, sequential analysis, and
finite population sampling.  These topics are important for a variety of
astronomical problems, arguing again that work on fundamentals will have
practical consequences for astronomers.

The {\it ISBA Bulletin} from the International Society for Bayesian Analysis,
available at \url{http://bayesian.org/}, is a good source for occasional
informal interchanges on these issues.  In a recent issue, ISBA President
Michael Jordan polled a number of leading statisticians (including some whose
work is largely frequentist) on what they thought were the principal open
problems in Bayesian statistics (Jordan 2011).  They noted that Bayesian and
frequentist methods can differ considerably on how to address model
selection, model misspecification, and model validation.  Computation is often
seen as difficult; approximate Bayesian computation (ABC) methods, as
described to us by Chad Schafer at this conference, may be an important
emerging approach.  The relationships between frequentist and Bayesian methods
need to be elucidated, such as connections between empirical Bayes and the
bootstrap and false discovery rate (FDR) control.  Choice of priors continues
to be an important issue. Concern was expressed about nonparametric and
semiparametric inference where it presently seems safer and easier to use
frequentist rather than Bayesian methods; this was discussed by Christopher
Genovese earlier in the conference.  In all of these areas, clarifying
foundations will directly affect practice.  And many of them are clearly
relevant to current and emerging astrostatistics problems.

\section{Multilevel models and multiple testing}

Let me elaborate on one item in Jordan's list as an example of where some
struggle at the Bayes/frequentist divide by statisticians and astronomers
together might pay dividends:  the role of multilevel modeling (empirical or
hierarchical Bayes) in multiple testing, where FDR control has become the
standard frequentist technique.  Statistical research in this area is
important for addressing challenges being raised by the astronomy data deluge
described earlier in this panel discussion by David van~Dyk.  The deluge
coming from synoptic surveys does not just provide astronomers {\em more} data
than we are used to; it also provides a {\em different kind} of data:
collections of modest-sized datasets (such as sparse, irregularly-sampled
light curves) for vast numbers of related objects.  Astronomers need methods
that can accurately and optimally accumulate information, not only within the
dataset for a particular object, but also across a population of related
objects.



This problem is not unique to astronomy.  It is arising in many disciplines,
motivating much current statistics research.  This research was the main theme
of a recent article by Bradley Efron entitled ``The future of indirect
evidence'' in the excellent cross-disciplinary journal {\it Statistical
Science} (Efron 2010).  Whereas conventional statistical methods accumulate
information about an object or process by repeated observations of the same
object or process, new data modalities require the ability to pool information
across ensembles of related objects or processes---``indirect evidence.'' 
Efron advocates empirical Bayes methods as a promising paradigm for using
indirect evidence, and false discovery rate control for the class of
problems where the goal is separation of a large ensemble of related
observations into discoveries and ``nulls.'' Efron's paper was published with
discussion; none of the discussants liked FDR, and neither do I.  For
astronomers, a catalog is not just a report of final classifications of
candidate sources.  Rather, it is a starting point for further analysis and
discovery, perhaps the most common goal being estimating population
distributions.  Catalogs produced by FDR control are ill-suited to this.

\begin{figure}[t]
\includegraphics[width=0.5\textwidth]{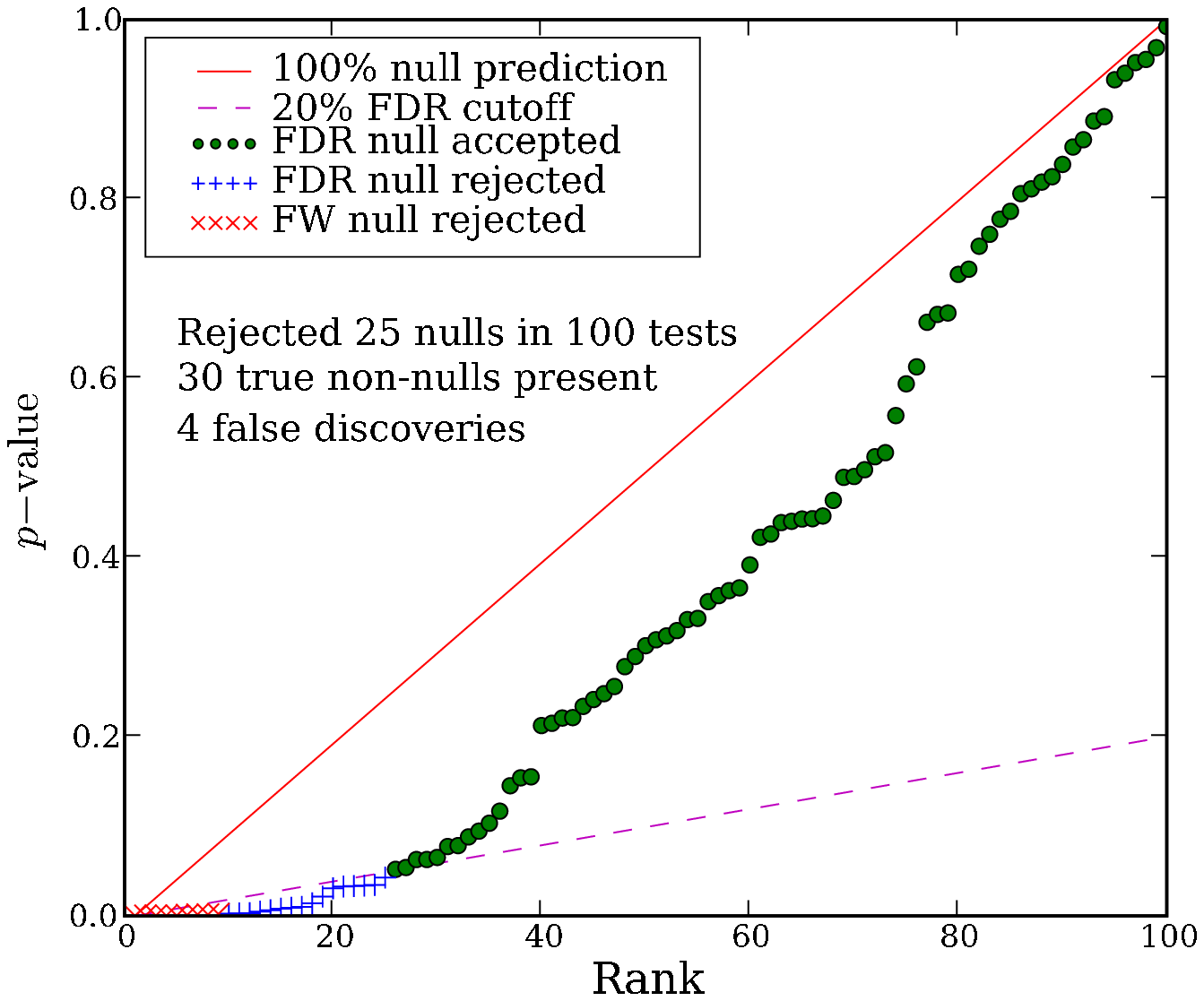}
\includegraphics[width=0.5\textwidth]{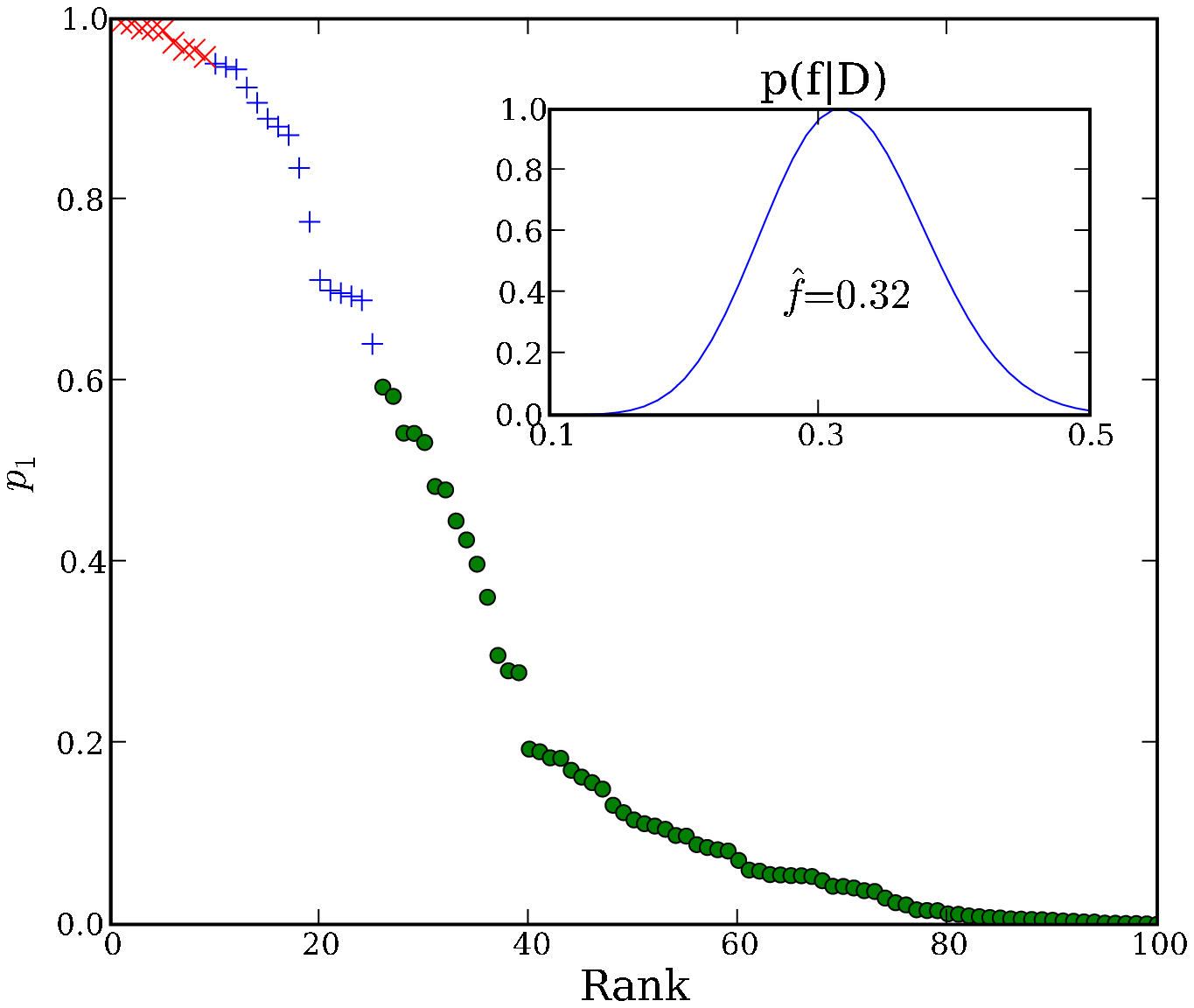} 
\caption{Detecting sources from noisy observations in a simple simulation.
{\em Left}: Discoveries using $p$-values with a threshold controlling FWER at 20\% (red Xs), or controlling FDR at 20\% (red Xs and blue crosses).
{\em Right}: Results from a hierarchical Bayesian analysis.  Points show marginal probabilities each observation is from a source; inset shows marginal density for the fraction, $f$, of observations that are sources.  See text for details.}
\label{panel_loredo_FDR.fig}
\end{figure}

Consider a simple hypothetical example.  Suppose an astronomer observes 100
candidate source locations (say, in a search for counterparts in a new
waveband), 30\% of which have emission; the fraction of emitting sources is
unknown to the astronomer (and one of the targets of study).  Each observation
is noisy and the sources are not strong, so not all emitting sources will be
securely detected.  A common goal is to present the community with a catalog
of detected sources which subsequently will be analyzed for various scientific
purposes.  In astronomy we traditionally set a threshold for catalog
membership, chosen to control the family-wise error rate (FWER) for a set of
tests of the null hypothesis that no emission is present---we pick a small
critical $p$-value that makes us confident that there are {\em no} false
sources in our catalog.  The virtue of this approach is that the final catalog
is essentially pure and can be studied in simple ways.  The left panel of
Figure~\ref{panel_loredo_FDR.fig} illustrates this, with 100 simulated
observations where the 30 sources have fluxes corresponding to a
signal-to-noise ratio $\textrm{SNR} = 2.2$.  The plotted points show the
$p$-values in rank order (effectively a cumulative histogram on its side).  To
control the FWER at 20\% (say), the $p$-value threshold is set to
$0.2/100=0.002$.  The red `X' symbols show the source detections that pass
this threshold (i.e., that fall below a horizontal line at $p=0.002$); only 10
of the 30 sources are detected, but there are no false discoveries.

The motivation for FDR control is that FWER control probably throws out many
real sources (it certainly does here).   The Benjamini-Hochberg (BH) method
for FDR control relaxes the acceptance criterion to be less conservative, in a
way that adapts to the evidence in the data.  To motivate it, the red line
shows the expected $p$-value vs.\ rank if only noise were present (so rank $=$
$100\times p$-value).  Note that the observed $p$-value distribution drops
well below the line; there is an excess of small values corresponding to
observations of real sources.  The dip is an indication that the null
hypothesis (all observations are from noise) is not relevant; we should be
able to somehow use the actually observed sample of $p$-values to set a
threshold.  The BH method does this by tilting the null line by an amount
determined by the target FDR (to the dashed purple line), establishing a
threshold at its intersection with the observed $p$-value distribution; Miller
et al. (2001) introduced this remarkably clever yet simple technique to
astronomers.  The blue crosses show the additional discoveries that result
from using BH to target an expected FDR of 20\%.  Now 25 sources are detected,
4 of which are false detections of noise.  Were the data real instead of
simulated (with known ground truth), we would guess that 20 of the 25
``discoveries'' are real sources, given the targeted FDR.

The problem is that the false discoveries pile up at the low $p$-values.  For
this simulation, the SNR was the same for all sources, but in realistic
settings the SNR will be lower for dim sources than for bright ones, so the
low $p$-values will tend to come from dim sources.  Applying the BH method to
this situation will give progressively greater pollution at dimmer fluxes. 
Simply knowing that you have controlled the FDR at some specified level for
the whole catalog does not help you accurately infer the run of $\log
N$--$\log S$ (log source counts {\em vs.} log flux, i.e., the number-size
distribution) or other interesting population-level quantities from the
catalog.  So FDR control addresses a particular question in an almost
miraculously beautiful way---nonparametrically, adaptively, and robustly---but
it does not provide results that let astronomers answer further, related
questions we want to address with the data.

The right panel of Figure~\ref{panel_loredo_FDR.fig} shows results from a
Bayesian multilevel model approach that attempts a soft (probabilistic)
classification.  It calculates a joint distribution for the classification of
each object (real source or noise) and the fraction, $f$, of the 100
observations that are true sources (the calculation uses a flat prior on $f$).
 The plotted points show the marginal probability, $p_1$, that each
observation belongs to the class of sources with nonzero emission.  The inset
shows the marginal posterior density for $f$.  It peaks at $\hat f = 0.32$,
near the true value of 30\%.  The calculation can accurately estimate the
fraction without specifying which specific observations are of actual sources.
 Many observations are very probably sources or nulls ($p_1 \approx 1$ or 0),
but there is a significant middle group with ambiguous classifications.  In
this group, we cannot be sure any particular observation is a real source; but
we can be sure that there are several real ones among those candidates, and
this lets us estimate $f$ accurately.  This kind of calculation can be used to
construct reliable $\log N$--$\log S$ curves and to make other inferences
without requiring hard thresholding (though various practicalities will likely
require some low threshold in most applications).  The method is not without
danger; e.g., setting the ``upper level'' prior (here, on $f$) needs some care
in more complex settings (e.g., Scott \& Berger 2006).

Statisticians themselves are not uniformly enthusiastic about FDR control. 
Gelman (2010) wrote: ``To me, the false discovery rate is the latest
flavor-of-the-month attempt to make the Bayesian omelette without breaking the
Bayesian eggs\ldots it can work fine if the implicit prior is ok\ldots but I
really don't like it as an underlying principle.''  The frequentist literature
on multiple testing itself recognizes that FDR control may not address the
science questions of interest in a particular study.  It includes alternatives
to FDR control, such as estimation of confidence bounds on the source fraction
advocated by Meinshausen \& Rice (2006) for some applications.  Tighter
interaction between astronomers and statisticians is needed to work out how
frequentist and Bayesian approaches to multiple testing might interact to
produce tools meeting astronomers' needs.  For example, can we simultaneously
have the robustness offered by BH FDR control and the soft thresholding
offered by Bayesian multilevel models, enabling a variety of subsequent
scientific analyses using the source detection results?

\section{Statistical analysis and the chain of discovery}

\begin{figure}[t]
\centering
\includegraphics[width=1.0\textwidth]{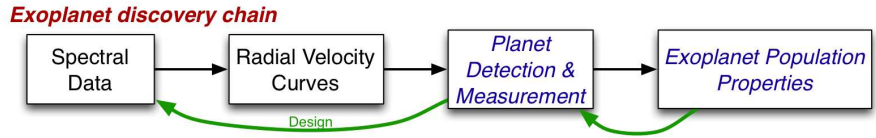}
\caption{Diagram of a discovery chain whereby exoplanets are found from periodic Doppler shifts in the spectra of the host stars.  Progress in detection and characterization of individual planets leads to studies of exoplanet populations, and improved design of the observational experiments and spectral analysis procedures.}
\label{panel_loredo_chain.fig}
\end{figure}

Frequentist methods tend to frame a data analysis task as a monolithic
decision, as if addressing that one decision were the sole goal of
data taking.  Indeed, this is made explicit in the decision-theoretic
formulation of frequentist estimation and testing.  But astronomers are seldom
seeking to produce a single terminal decision from their data.  Instead our
observing and cataloging and  modeling are all just steps in what one might
call {\em unfolding chains of discovery.}  An astronomical problem is often
first tackled with sequential experimentation and exploration, starting a
chain of discovery leading from study of individual objects to study of
populations.  Figure~\ref{panel_loredo_chain.fig} diagrams an example of such
a chain for extrasolar planet science using radial velocity data, where
planets orbiting other stars are detected from time-dependent Doppler shifts
of the spectra of their host stars.  Each of the black arrows represents a
complicated data analysis problem, converting spectral data into radial
velocity curves, modeling these curves to detect planets (as Philip Gregory
described at this meeting), and inferring properties of exoplanet populations
from the individual planetary measurements.  But effective analysis, and even
effective data acquisition, requires knowledge from the later steps, so a
feedback loop is established.  We need a broad statistical approach that
facilitates building such chains (and loops) of discovery.

This notion of a discovery chain is related to the type of problem studied in
the branch of statistics known as sequential analysis.  A pioneer of this
area, Herman Chernoff, has an intriguing perspective on its relevance to the
scientific process more generally (the following quote is from an
interview with Chernoff reported in Bather 1996):
\begin{quote}
I became interested in the notion of experimental design in a much broader
context, namely: what's the nature of scientific inference and how do people
do science?  The thought was not all that unique that it is a sequential
procedure\ldots.  Although I regard myself as a non-Bayesian, I feel in
sequential problems it is rather dangerous to play around with non-Bayesian
procedures\ldots.  Optimality is, of course, implicit in the Bayesian
approach.
\end{quote}
An important direction for future fundamental work in statistics would be
explicit recognition that most scientific data analysis tasks are just steps
in an ongoing sequence of analyses---an unfolding chain of discovery.  Efron's
``indirect evidence'' is a special case of this, where one seeks a framework
that can integrate inference about individuals with inference about
populations.  Given Chernoff's remarks, it is perhaps not surprising that
Bayesian ideas are playing an important role in working out how to use
indirect evidence, via empirical and hierarchical Bayes methods.  I suspect
the future of statistics will involve a more thorough integration of Bayesian
ideas into statistical practice, if only to enable development of even more
elaborate discovery chains.  I anticipate that statistical challenges in
modern astronomy will be both drivers and beneficiaries of such developments.

\bigskip

\noindent{\large\bf References}

\medskip

\hi Bather, J. (1996) A Conversation with Herman Chernoff,
{\it Statist. Science}, 11, 335--350

\hi Bayarri, M. J. \& Berger, J. O. (2004) The interplay of Bayesian and
frequentist analysis, {\it Statist. Science}, 19, 58--80

\hi Efron, B. (2005) Bayesians, frequentists, and scientists, {\it J. Am.
Stat. Assoc.}, 100, 1--5

\hi Efron, B. (2010) The future of indirect evidence, {\it Statistical
Science}, 25, 145--157

\hi Gelman, A. (2010) Bayesian statistics then and now (discussion of Efron's
``The future of indirect evidence''), {\it Statistical Science}, 25,
162--165

\hi Jordan, M. (2011) What are the open problems in Bayesian statistics?, {\it
The ISBA Bulletin}, 11(1), 1--4 

\hi Little, R. (2006) Calibrated Bayes: a Bayes/frequentist roadmap, {\it
Amer. Statist.}, 60, 213--223

\hi Meinshausen, N. \& Rice, J. (2006) Estimating the proportion of false null
hypotheses among a large number of independently tested hypotheses, {\it
Annals Statistics}, 34, 373--393

\hi Miller, C.~J., Genovese, C., Nichol, R.~C., et al.\  (2001) Controlling
the False-Discovery Rate in astrophysical data analysis, {\it Astron. J.},
122, 3492-3505

\hi Scott, J. G. \& Berger, J. O. (2006) An exploration of aspects of Bayesian
multiple testing, {\it  J. Statist. Plann. Inference}, 136, 2144�2162.

\end{document}